\documentclass[aps,prl,secnumarabic,twocolumn,superscriptaddress,a4paper,showpacs,floatfix]{revtex4-1}
\usepackage{amsfonts,amssymb,amsmath}
\usepackage[utf8]{inputenc}

\usepackage[usenames,dvipsnames]{color}
\usepackage[pdftex,breaklinks=true,colorlinks=true]{hyperref}
\usepackage[pdftex]{graphicx}
\usepackage{pict2e}
\usepackage[caption=false]{subfig}
\usepackage{epstopdf}
\usepackage{filecontents}

\newcommand{\La}{\line (1,0  ){12}}
\newcommand{\Lb}{\line (3,5 ){6}}
\newcommand{\Lc}{\line (-3,5 ){6}}
\newcommand{\Ld}{\line (-1,0){12}}
\newcommand{\Le}{\line (-3,-5){6}}
\newcommand{\Lf} {\line(3,-5){6}}
\newcommand{\C} {\circle*{4}}

\newcommand{\pA}{\put(-6,-10)}
\newcommand{\pB}{\put(6,-10)}
\newcommand{\pC}{\put(12,0)}
\newcommand{\pD}{\put(6,10)}
\newcommand{\pE}{\put(-6,10)}
\newcommand{\pF}{\put(-12,0)}

\newcommand{\Hex}{\pA{\C}\pB{\C}\pC{\C}\pD{\C}\pE{\C}\pF{\C}}

\newcommand{\ha}{
  \begin{picture}(32,18)(-16,-5)
    \Hex\pA{\La}\pC{\Lc}\pE{\Le}
  \end{picture}
}
\newcommand{\hb}{
  \begin{picture}(32,18)(-16,-5)
    \Hex
    \pB{\Lb}\pD{\Ld}\pF{\Lf}
  \end{picture}
}

\begin{document}

\title{Phase Diagram of an Extended Quantum Dimer Model on the Hexagonal Lattice}

\author{Thiago Schlittler}
\email{thiago.schlittler@upmc.fr}
\affiliation{Laboratoire de Physique Th\'eorique de la Mati\`ere Condens\'ee, CNRS UMR 7600,
Universit\'e Pierre et Marie Curie, Sorbonne Universit\'es, 4 Place Jussieu, 75252 Paris Cedex 05, France}
\author{Thomas Barthel}
\homepage{http://www.manyparticle.org/~barthel}
\affiliation{
Laboratoire de Physique Th\'{e}orique et Mod\`{e}les Statistiques, 
Universit\'{e} Paris-Sud, CNRS UMR 8626, 91405 Orsay Cedex, France}
\affiliation{Department of Physics, Duke University, Durham, North Carolina 27708, USA}
\author{Gr\'egoire Misguich}
\email{gregoire.misguich@cea.fr}
\affiliation{Institut de Physique Th\'eorique, Universit\'e Paris Saclay, CEA, CNRS, F-91191 Gif-sur-Yvette, France}
\author{Julien Vidal}
\email{vidal@lptmc.jussieu.fr}
\affiliation{Laboratoire de Physique Th\'eorique de la Mati\`ere Condens\'ee, CNRS UMR 7600,
Universit\'e Pierre et Marie Curie, Sorbonne Universit\'es, 4 Place Jussieu, 75252 Paris Cedex 05, France}
\author{R\'emy Mosseri}
\email{remy.mosseri@upmc.fr}
\affiliation{Laboratoire de Physique Th\'eorique de la Mati\`ere Condens\'ee, CNRS UMR 7600,
Universit\'e Pierre et Marie Curie, Sorbonne Universit\'es, 4 Place Jussieu, 75252 Paris Cedex 05, France}

\begin{abstract}
We introduce a  quantum dimer model on the hexagonal lattice that, in addition to the standard
three-dimer kinetic and potential terms,
includes a competing potential part
counting dimer-free hexagons.
The zero-temperature phase diagram is studied by means of quantum Monte Carlo simulations, supplemented by variational arguments. It reveals 
some new crystalline phases and a cascade of transitions with rapidly changing flux (tilt in the height language). 
We analyze perturbatively the vicinity of the Rokhsar-Kivelson point, showing that this model has
the microscopic ingredients needed for the ``devil's staircase'' scenario [E. Fradkin {\it et al.} Phys. Rev. B {\bf 69}, 224415 (2004)],
and is therefore expected to produce fractal variations of the ground-state flux.
\end{abstract}

\maketitle

The study of hard-core dimer coverings has a long history. From the mapping to Pfaffians and determinants by Kasteleyn \cite{kasteleyn_statistics_1961, fisher_statistical_1961}, the solution of two-dimensional Ising models \cite{fisher_dimer_1966}, the height representation and its continuum limit \cite{blote_roughening_1982}, or the connection to the Coulomb gas and conformal field theory \cite{alet_interacting_2005,papanikolaou_quantum_2007}, dimer models have found numerous applications in various fields of statistical physics. Motivated by the physics of resonating valence bond systems, Rokhsar and Kivelson (RK) \cite{rokhsar_superconductivity_1988} added quantum dynamics to the dimer model,
leading to the so-called quantum dimer model (QDM), which later led to  tractable models with rich phase diagrams closely related to lattice gauge theories \cite{fradkin_short_1990}.
Importantly, QDMs appeared in different contexts when describing the dynamics in a constrained low-energy manifold, such as in frustrated Ising models in weak transverse fields \cite{moessner_ising_2001}.
QDMs also gained a new dimension with the discovery of liquid phases with topological order in nonbipartite lattices \cite{moessner_resonating_2001,misguich_quantum_2002}, where they shed some light on the long-sought resonating valence bond liquids. This field also benefited from recent progress in making quantitative connections between spin-$1/2$ Heisenberg magnets
with quantum disordered ground states and QDMs \cite{poilblanc_effective_2010,rousochatzakis_quantum_2014}.

In most QDMs studied so far, a kinetic term (associated with on-plaquette dimer flips) competes with a diagonal term  proportional to the number of such ``flippable'' plaquettes.
When the kinetic and the potential terms are equal at the so-called RK point, the ground states are exactly known \cite{rokhsar_superconductivity_1988}.
In the height language, appropriate for bipartite lattices, such a RK point corresponds to a transition from a ``flat'' phase to a maximal slope phase~\footnote{See Ref.~\cite{banerjee_interfaces_2014} and references therein for the latest numerical results on the square lattice QDM phase diagram.}. A richer behavior is however expected near that point for more generic interactions between dimers \cite{fradkin_bipartite_2004,vishwanath_quantum_2004}. In particular, within a field theoretic approach, a devil’s staircase of commensurate and incommensurate phases is predicted \cite{fradkin_bipartite_2004,vishwanath_quantum_2004,levitov_equivalence_1990}, corresponding to a fractal tilt variation as a function of the Hamiltonian parameters.

In this Letter, we show that a natural generalization of the hexagonal lattice QDM \cite{moessner_phase_2001,schlittler_phase_2015} provides a microscopic model with this phase structure. We analyze the two-parameter phase diagram spanned by the standard potential term counting flippable plaquettes and another term counting dimer-free plaquettes.
The model is studied perturbatively near the RK point and with quantum Monte Carlo (QMC) simulations elsewhere, supplemented by variational arguments. We observe a sequence of closely spaced phase transitions with a gradual change of the flux density and crystalline structures with strongly varying unit cell sizes in agreement with the scenario of Refs.~\cite{fradkin_bipartite_2004,vishwanath_quantum_2004}.

\emph{Model.---}Let us consider a QDM  with the standard kinetic term and four potential terms:
\begin{eqnarray}
  \hat{H}=&-t&\sum_h \left(
  \left|\ha\right>\left<\hb\right| +{\rm H.c.}
  \right) + \sum_{j=0}^3 v_j \hat{n}_j,
\label{eq:Hgen}
\end{eqnarray}
where the operator $\hat n_j$ counts the total number of hexagonal plaquettes with $j$ dimers (called a \mbox{$j$-plaquette}). 
Because of the two sum rules \cite{schlittler_phase_2015,SuppMat} $\hat n_0+\hat n_1+\hat n_2 + \hat n_3=N$ and $ 2\hat n_0+\hat n_1-\hat n_3=0$, these potential terms are not independent and we hence choose to keep only $\hat n_0$ and $\hat n_3$. Also, we denote densities $\rho_j= \langle \hat n_j\rangle /N$ in the form $\vec{\rho}=(\rho_0,\rho_1,\rho_2,\rho_3)$ and fix $t=1$, unless specified differently.
The model studied by Moessner \emph{et al.} \cite{moessner_phase_2001} has $v_0=0$, while the two models $(v_0=\pm1,v_3=0)$ are relevant for  Ising string nets \cite{schulz_ising_2014}. We study rectangular clusters with periodic boundary conditions and $N=L_x\times L_y$ hexagonal plaquettes.

Our analysis relies on the notion of flux:
dimer coverings can be grouped into topological sectors \cite{SuppMat} labeled by two integer fluxes $(F_x,F_y)$, which
are invariant under local dimer moves.
As discussed below, 
for ground states, one of the two fluxes is zero and we can restrict 
ourselves to $F_x=0$ and work with  $f:=F_y/L_y\geq 0$.

\begin{figure}[t]
 \includegraphics[width=0.95\linewidth]{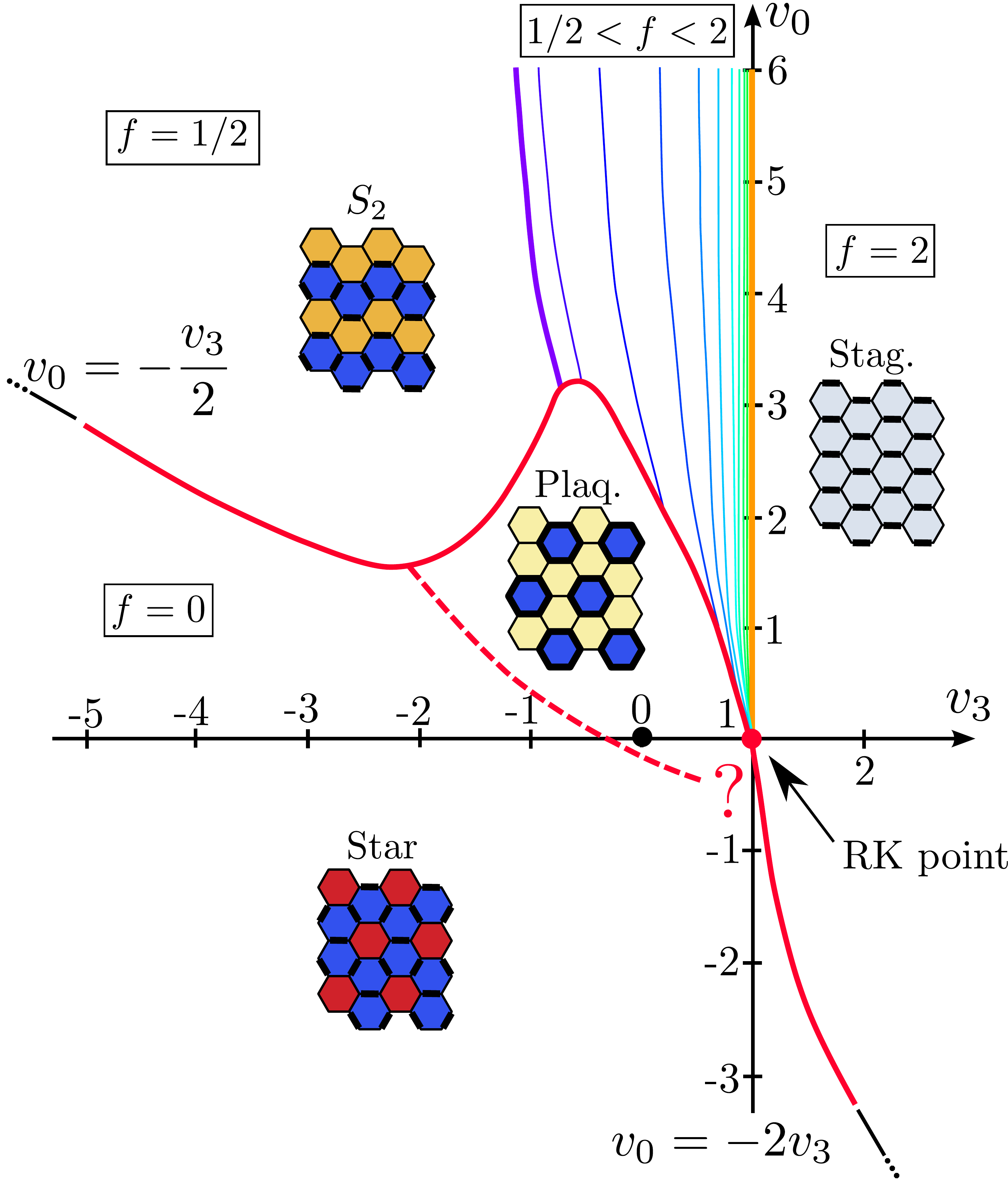}
 \caption{Schematic phase diagram from QMC simulations ($L_x=L_y=60$). The $(v_0,v_3)$ plane is divided into five regions: a staggered phase with the maximal flux ($f=2$), the star
 and the plaquette phases ($f=0$),  the $S_2$ phase ($f=1/2$), and the fan region, containing a cascade of flux sectors $1/2\leq f<2$.
 The plaquette color indicates the dimer density (same scale as Figs.~\ref{fig:SHF} and ~\ref{fig:variable}).}  
\label{fig:phase_diagram}
\end{figure}

\emph{Classical limit.---}Let us consider the classical limit $t=0$. Setting \mbox{$v_0=\sin\alpha$}, $v_3=\cos\alpha$, and defining $\alpha_1=\arctan(-2)$, $\alpha_2=\pi/2-\alpha_1$, one finds 
three crystals as $\alpha$ is varied: (i) for $\alpha\in\left[\pi/2,\alpha_1\right]$, the threefold degenerate staggered crystals (nonflippable configurations) with maximum flux $f=2$, vanishing energy, and $\vec{\rho}=(0,0,1,0)$, (ii) for $\alpha\in\left[\alpha_1,\alpha_2\right]$, the (threefold degenerate) star crystal in the $f=0$ sector (Fig.~\ref{fig:phase_diagram}) with $\vec{\rho}=(1/3,0,0,2/3)$, (iii) for $\alpha\in\left[\alpha_2,\pi/2\right]$, a 12-fold degenerate crystal~\footnote{The factor 12 is due to three equivalent orientations for the chains of 3-plaquettes, two translations perpendicular to the chains, and the period of 2 along the chains.} denoted $S_2$, within the  $f=1/2$ sector, with $\vec{\rho}=(0,1/2,0,1/2)$.
The point $\alpha=\pi/2$ is highly degenerate, since any configuration without \mbox{$0$-plaquettes} is a ground state, and such states exist in all flux sectors. This  degeneracy is lifted when $t\ne 0$, leading to a nontrivial ground-state flux variation  as discussed below.
 
\emph{Phase diagram.---}We studied the phase diagram with  QMC simulations
using the  mapping to an Ising-type model described in Refs.~\cite{moessner_phase_2001,schlittler_phase_2015,SuppMat}.
Specifically, results displayed in Fig.~\ref{fig:phase_diagram}
have been obtained for a torus with $60\times60$ plaquettes,
flux sectors $f=0,\frac{1}{10},\frac{2}{10},\dots,2$, inverse temperature 
$\beta=9.6$, and imaginary-time step $\Delta\beta=0.01$.

\begin{figure}
 \includegraphics [width=0.9\linewidth]{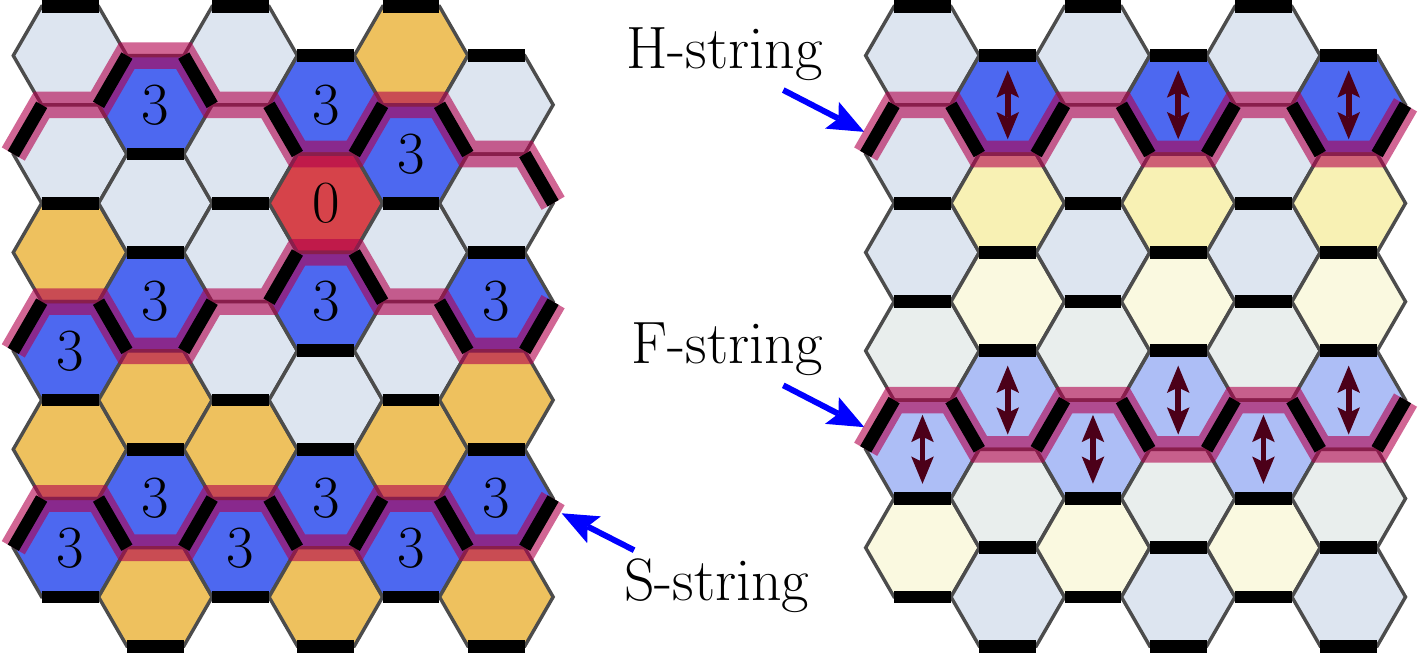}
 \caption{Left: a configuration of three strings and the corresponding dimer covering, with $0$- and $3$-plaquettes.
 Left bottom and right:
 three variational classes of dynamically constrained strings, called 
$S$-, $H$-, and $F$-strings. $S$-strings are in a static zigzag configuration, $H$-strings ($F$-strings) are allowed to fluctuate by one row in every 
second column (in every column). Arrows indicate the fluctuations of the strings, each corresponding to a 3-plaquette flip.
Dimer densities are indicated according to the color scale of Fig.~\ref{fig:variable}.
For $H$-  and $F$-strings, the shown dimer densities correspond to a 
superposition of the allowed configurations.
}
 \label{fig:SHF}
\end{figure}

\emph{$\bullet$ $f=2$}. In this region, ground states are isolated staggered configurations with vanishing energy.
The Hamiltonian is positive definite in the upper right quadrant, and the $f=2$ region also extends to a large part of the lower right quadrant, down to the  boundary with the $f=0$ sector.

\emph{$\bullet$ $f=0$}.
The star and plaquette crystals found in this region also exist in the $v_3$-only model \cite{moessner_phase_2001,schlittler_phase_2015}
and are separated by a first-order transition (dashed line). 
The star phase is adiabatically connected to the (threefold degenerate) crystalline configurations found for $t=0$. 
The latter simultaneously maximize the number of 3- and 0-plaquettes, and the star phase thus fills a large part of the $(v_3<0,v_0<0)$-quadrant and also extends into the neighboring quadrants. 
On the $v_0=0$ line, the star phase gives way to the plaquette phase through a first-order transition
at $v_3 = -0.228(2)$ \cite{moessner_phase_2001,schlittler_phase_2015}. The plaquette phase is defined
by continuity with the ``ideal'' plaquette state, which is an uncorrelated product of resonating $3$-plaquettes $|\ha\rangle+|\hb\rangle$.
In the vicinity of the RK point, as is already the case for $\hat{H}(t,v_0=0,v_3)$ \cite{schlittler_phase_2015}, the large (diverging) correlation length makes it difficult to discriminate numerically between the star and plaquette phases,
hence the question mark in Fig.~\ref{fig:phase_diagram}. This phenomenon is likely to be related to the $U(1)$ regime observed in the square lattice QDM \cite{banerjee_interfaces_2014}.

\emph{$\bullet$ $f=1/2$}.
In most of this region, the system forms a 12-fold degenerate crystalline phase, adiabatically connected to the $S_2$ configuration.

\emph{$\bullet$ $1/2<f<2$}. This is the most interesting part of the phase diagram, which we call the fan region.
To understand  the flux variations taking place there,
we recall that any dimer configuration can be represented equivalently as a  
configuration of nonintersecting strings on the hexagonal lattice 
\cite{SuppMat}. For $F_x=0$, these are $N_s=(2L_y-F_y)/3$ closed loops along 
the toroidal $x$ direction of the lattice. Starting from the staggered dimer 
covering ($f=2$) displayed in Fig.~\ref{fig:phase_diagram}, on each string 
path, empty and covered edges alternate.
The corresponding dimer covering is obtained by doing so-called loop updates, i.e., exchanging empty and covered edges along the string 
paths. Each string reduces the flux $F_y$ by three units.
In reverse, starting from an arbitrary configuration, the strings correspond to paths
where dimer-free horizontal edges alternate with dimers on tilted edges (see Fig.~\ref{fig:SHF}).
The number of $3$-plaquettes along a string is maximized if it runs parallel to one of the three edge orientations of the lattice.
This is why, for $v_3<1$, strings are on average parallel to one of the edge orientations
and why ground states are found in sectors with one vanishing flux quantum number
($F_x$ vanishes for strings winding in the $x$ direction only).
Strings can reduce their kinetic energy by oscillating in the perpendicular direction, limited by the string noncrossing condition 
and by avoidance of $0$-plaquettes for large $v_0$ (see Fig.~\ref{fig:SHF}).

When $v_3$ is decreased below $1$, the staggered configuration is destabilized by  string insertion. At low string densities ($f$ slightly below 2) strings are far apart and strongly delocalized.
A reduction of $v_3$ causes an increase of $\rho_3$, which is realized
through a
higher string density (decrease of the flux) and ``stiffer'' strings (reduced lateral motion).
Each time a new string is added upon decreasing $v_3$, the increased $\rho_3$ compensates the energy cost associated with the higher degree of 
localization.
When increasing $v_0$ for a fixed $v_3<1$, configurations with more 0-plaquettes become less favorable such that string delocalization gets more 
restricted. At certain transition points, it becomes favorable to remove a 
string (flux increase), freeing some space for other strings to fluctuate more freely.
When $\rho_0$ becomes negligible, a further increase of $v_0$ has no effect. This regime, where the isoflux lines become parallel,
is equivalent to perturbing the (degenerate) classical point $(t, v_0,v_3 ) = (0,1,0)$ with a weak $t$ and $v_3$, where a fan like phase diagram similar to that described in Ref.~\cite{papanikolaou_devils_2007} is expected.

\begin{figure}[t]
  \includegraphics [width=\columnwidth]{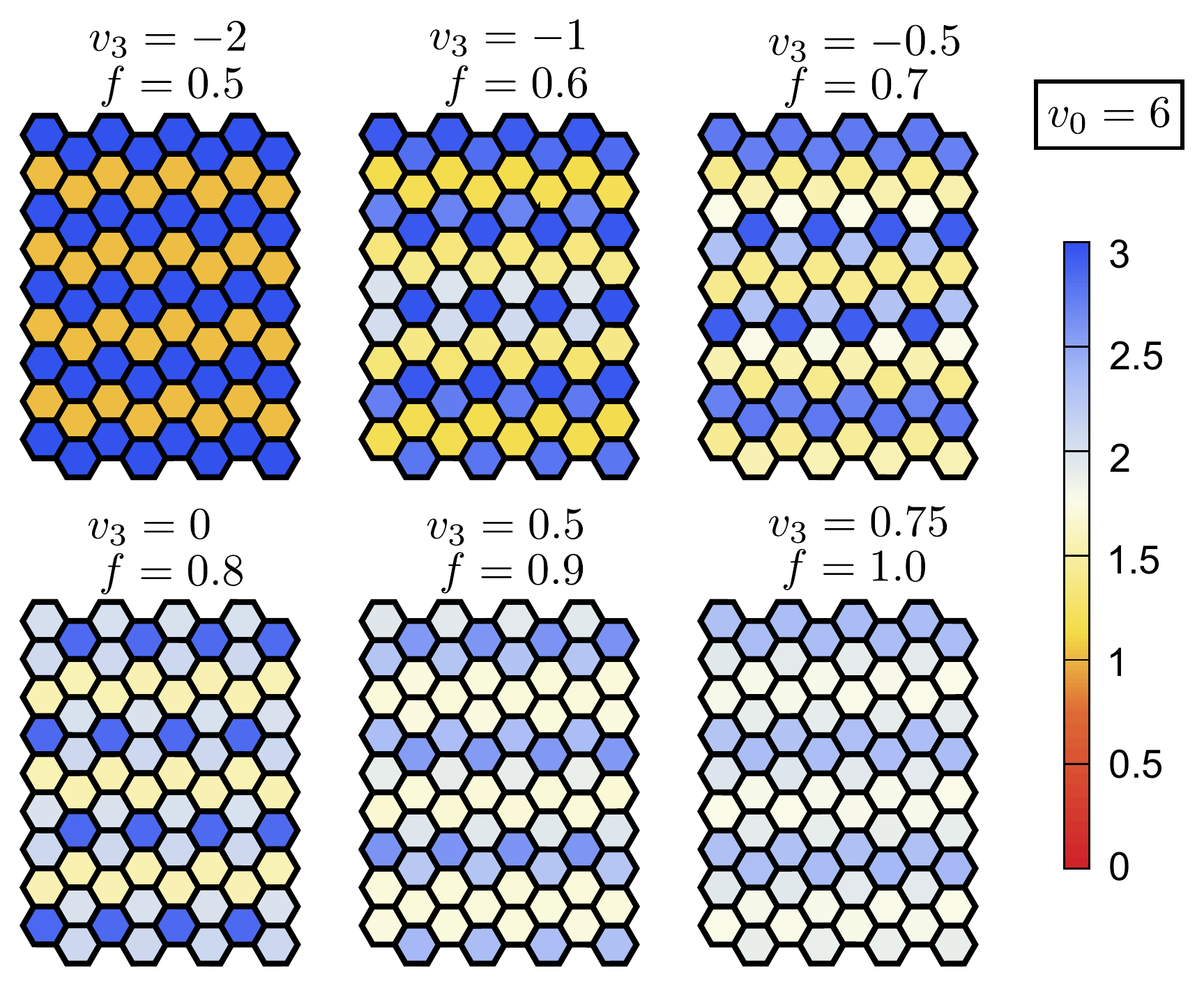}
 \caption{Dimer density per hexagon in the fan region, according to the color scale on the right.	}
 \label{fig:variable}
\end{figure}

For $f\lesssim 1$ the average interstring distance is sufficiently low that
the ground states are dominated by straight-string configurations. For generic fluxes, one expects
complex correlated string states (some are described in Ref.~\cite{SuppMat}), but simple spatial structures
involving horizontal chains of hexagons with higher densities of $3$-plaquettes are also observed in some low-flux parts of the fan (see Fig.~\ref{fig:variable}).
These can be qualitatively understood in terms of the following typical 
configurations of strings that are dynamically constrained by the presence of 
neighboring strings: ``$S$-strings'' are static zigzag configurations
(corresponding to zigzag arrangements of 3-plaquettes, energetically 
favored at large negative $v_3$). With respect to such a reference 
configuration, ``$H$-strings'' can fluctuate in every second column of hexagons, 
up and down by one row. ``$F$-strings'' are the most mobile among the three 
 classes, and are allowed to fluctuate up and down by one row in 
every column as indicated by arrows in Fig.~\ref{fig:SHF}.
At $f=0.8$ and $1$, for instance, we recognize periodic arrays of $H$- ($F$-) strings at distance $d = 2.5$ ($d = 3$)
\footnote{String distances $d$ are measured in units of the distance between hexagon centers.}, as shown in Fig.~\ref{fig:variable}.
Importantly, no $0$-plaquettes
are generated if the above strings have minimum interstring distances of
$d_{S-S}^{\rm min}=2$, $d_{H-H}^{\rm min}=2.5$, $d_{F-F}^{\rm min}=3$, and $d_{F-H}^{\rm min}=2.75$.
These building blocks are therefore appropriate to describe qualitatively the large-$v_0$ and $f\lesssim 1$ part of the fan \cite{SuppMat}.

Finally, simple variational arguments provide approximate expressions for the flux transition lines.
For example, one can compute  the energy change associated with the insertion of an $H$-string in a perfect $S_2$ crystal ($S$-strings at distance 2), which 
corresponds to an infinitesimal increase of the flux density (due to the different $d^{\rm min}$, five $S$-strings should be  replaced by four $H$-strings to keep the total system size constant \cite{SuppMat}). 
This yields $v_3=-1$ for the transition towards the fan region at large $v_0$, in reasonable agreement with the numerics.

As the interplay between $v_3$ and $v_0$ is especially complex for 
low $v_0$ (when $\rho_0$ is not negligible), we analyzed  the $v_3=0$ line with finer flux steps. Starting from very large $v_0$ the flux decreases (staying close to $f=0.8$) down to $v_0\simeq 2.4$ where it drops to $f=0$. This flux drop is a generic feature of the interface with the $f=0$ region.
Toward the RK point the ground-state flux sectors get pinched, a feature that we now discuss.

\begin{figure}[t]
 \includegraphics[width=0.95\linewidth]{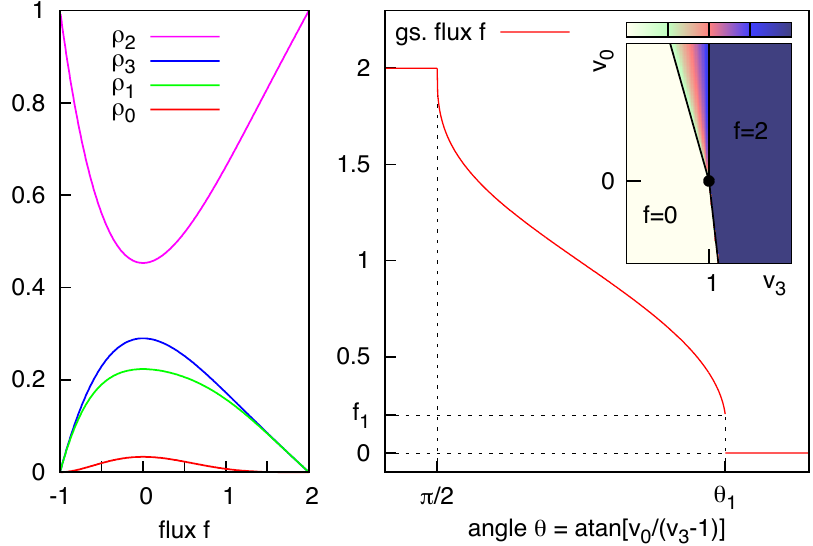}
 \caption{Perturbation theory near the RK point. Left: $j$-plaquette densities $\rho_j$ as functions of the flux density $f$. Right: ground-state flux density $f$ as a function of the angle $\theta$ that parametrizes the perturbation.  The nontrivial region lies between $\pi/2$ (transition out of the staggered phase) and $\theta_1\simeq 1.84695$, where $f$ drops discontinuously from $f_1\simeq 0.195654$ to $0$. The transition from $f=0$ to $f=2$ occurs at $\theta_2\simeq 4.8268$ (not shown).}
 \label{fig:f_theta_rho}
\end{figure}

\emph{Perturbative analysis.---}At the RK point,  the ground states of all flux sectors are degenerate, and are equal-amplitude superpositions of all dimer configurations in the corresponding sector. At first order in $v_0/t$ and $(v_3-1)/t$, the energy density in sector $f$ reads \mbox{$e(f)=v_0\rho_0(f)+(v_3-1)\rho_3(f)$}.
We compute the $j$-plaquette densities $\rho_j(f)$ as expectation values of the operators $\hat n_j$ (diagonal in the dimer basis) with respect to the unperturbed RK states, using an analytical transfer-matrix approach \cite{stephan_shannon_2009,SuppMat}.
Setting $v_0=\sin \theta$ and $v_3-1=\cos \theta$, we minimize $e(f)$ for each value of $\theta$ to obtain $f(\theta)$ as displayed in Fig.~\ref{fig:f_theta_rho}. A continuous variation of $f$ is found in the interval $\theta\in[\pi/2,\theta_1 \simeq 1.84695]$, which corresponds to the fan region in the phase diagram of Fig.~\ref{fig:phase_diagram}.
Interestingly, $f$ jumps discontinuously to zero at $\theta_1$. For $\theta\in[\theta_1,\theta_2 \simeq 4.8268 ]$, the ground state is in the $f=0$ flux sector, and it jumps to $f=2$ for $\theta\in[\theta_2,\pi/2]$. Note that, at this order, wave functions remain RK states, which are translation-invariant dimer liquids with algebraic correlations (for $f<2$). 

\emph{Field theory.---}To connect our perturbative and numerical results concerning the flux variations, let us turn to the height representation \cite{youngblood_correlations_1980,blote_roughening_1982,nienhuis_triangular_1984,levitov_equivalence_1990,henley_relaxation_1997}. Dimer coverings are mapped to membranes embedded in a cubic lattice, whose average tilt is directly related to the flux \footnote{A vanishing flux corresponds to membranes being, on average, normal to the $(1,1,1)$ direction.}. In this language the QDM becomes a quantum roughening problem \citep{levitov_equivalence_1990}. Long-distance properties are captured by taking the continuum limit of the height model and, in our case, the RK point is described by a massless Gaussian field theory \cite{henley_relaxation_1997}.
Fradkin \emph{et al.} \cite{fradkin_bipartite_2004} and Vishwanath \emph{et al.} \cite{vishwanath_quantum_2004} discussed
how the action is modified in the presence of generic perturbations, through a renormalization group (RG) analysis \cite{Grinstein1981-23} predicting nonvanishing flux phases. 
A cubic interaction for the height, with three spatial derivatives, is the leading term favoring $f\neq0$.
In our problem we observe that $v_0$ induces a flux density perpendicular to some edges of the hexagonal lattice. This implies that the sign of the corresponding coupling is negative in the notation of Ref.~\cite{fradkin_bipartite_2004}. At this stage, the system would be gapless with a linear dispersion at small momenta.
However, the site positions and the microscopic heights are both discrete and form a 3D lattice $\mathcal{L}$. For the (coarse-grained) height field, potential terms
 that respect the symmetries of $\mathcal{L}$ will be generated upon  integration over the short-distance fluctuations.
They can be written as $V(h,\vec{r})=\sum_{{\bf K}=(K_0,\vec{K})\in \mathcal{L}^*}V_{{\bf K}}e^{i (K_0 h + \vec{K}\cdot\vec{r})}$,
where the sum runs over the reciprocal lattice vectors of $\mathcal{L}$.
When the average flux (tilt) is commensurate with the lattice, it corresponds
to some reciprocal lattice vector $\bf K$ and the associated locking term $V_{{\bf K}}$ is then asymptotically relevant in the RG \cite{fradkin_bipartite_2004}, leading to gapped crystals. However, as explained in Ref.~\cite{fradkin_bipartite_2004}, these gaps can become exponentially small in $1/f$ close to the RK point. Since crystals for rational fluxes with small denominators are more stable, their range of attraction in the RG is larger compared to others and, for the phase diagram close to the RK point, one thus expects  a fractal succession of commensurate phases --- a ``devil's staircase''. At the smaller fluxes, stronger quantum 
fluctuations can outweigh locking terms and impose irrational flux densities such that gapless incommensurate structures are possible.

\emph{Conclusion.---}The extended QDM \eqref{eq:Hgen} is the first candidate for a microscopic realization of the ``Cantor deconfinement'' scenario, which predicts that a fractal succession of flux sectors occurs near the RK point. Whether the flux varies continuously, in a fractal way, or assumes only a finite number of values \footnote{Note that at sudden jumps of the flux from $f'$ to $f''>f'$, as occurs in the numerical and perturbative phase diagrams of Figs.~\ref{fig:phase_diagram} and \ref{fig:f_theta_rho}, all fluxes in $[f',f'']$ are degenerate. In doubt, the corresponding ground states can always be realized as phase-separated states of fluxes $f'$ and $f''$.} is impossible to answer with QMC simulations. Indeed, although we can simulate large lattices, available flux sectors correspond to a small set of rational values. Additionally, intrasector gaps become very small near the RK point and render simulations difficult. However, the fact that all flux sectors 
for $1/2<f<2$ occur in the QMC results and the width variations of the corresponding regions in the phase diagram plead in favor of the realization of a fractal in the thermodynamic limit.

Finally, let us note that flux sequences found here cannot occur for square lattice models with single-plaquette Hamiltonians. In that case, the sum rule $n_0=n_2$ makes any QDM with potential terms $\sum_j v_j \hat{n}_j$ equivalent to the original RK model, which lacks intermediate-flux phases.

\emph{Acknowledgments.---}We wish to thank S. Capponi for useful discussions.
G.~M. is supported by a JCJC grant from the Agence Nationale pour la Recherche (Project No. ANR-12-JS04-0010-01).

\bibliography{hexQDM,specific}{}
 
\onecolumngrid
\newpage
\twocolumngrid

\appendix
\label{appendix}

\begin{center}
\textbf{\large Supplemental Material for
``Phase Diagram of an Extended Quantum Dimer Model on the Hexagonal Lattice''}
\end{center}

\newcounter{appxcnt}
\setcounter{appxcnt}{1}
\newcommand{\Section}[1]{\section{Appendix \Alph{appxcnt}:\hspace{1ex} #1}\stepcounter{appxcnt}}
\label{supp_mat}

\Section{Topological flux quantum numbers}
\begin{figure*}[t]
  \includegraphics [width=0.8\linewidth]{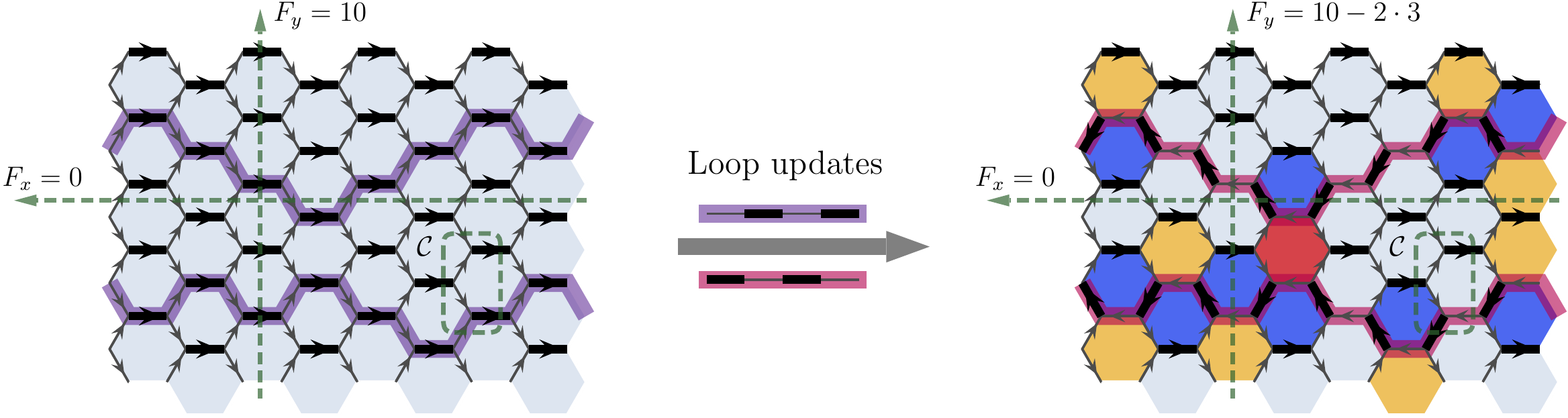}
  \hspace*{0.8cm}
  \includegraphics [width=0.027\linewidth]{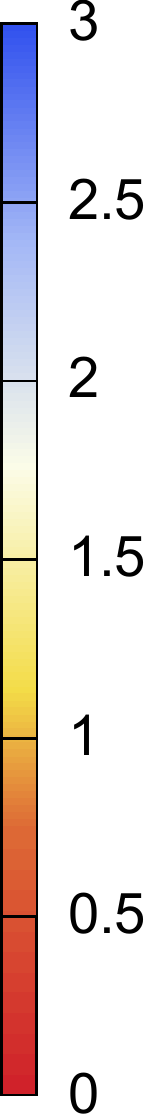}
 \caption{(Color online) Left: The perfectly staggered state ($f=2$) and a configuration of two strings on a torus with $8\times 5$ plaquettes. Right: The dimer covering that corresponds to the string configuration on the left is obtained by doing loop updates on the strings' paths, exchanging empty and occupied edges. Dimers carry two units of flux and empty edges one unit of flux in the indicated directions. Due to the dimer constraints, the flux through contractible loops such as $\mathcal{C}$ is always zero and the two nontrivial flux quantum numbers $F_x$ and $F_y$ for the 2D torus are invariant under local rearrangements. Only global loop updates, such as those indicated, change $F_x$ and $F_y$.
 The color of each plaquette indicates its dimer density, according to the scale shown on the right (same as in Fig.~3 in the main text).
 }
 \label{fig:stringsAndFlux}
\end{figure*}
The hexagonal lattice is bipartite: all the neighbors of a site in the ``even'' sublattice belong to the ``odd'' sublattice, and vice versa.
Let us recall that, with periodic boundary conditions, the set of all dimer coverings on a bipartite lattice
breaks into topological sectors which are stable under any local dimer move (including single-plaquette flips). These sectors can
be labeled by a pair of flux quantum numbers $(F_x, F_y)$. Locally, a ``magnetic'' field $\vec B$ can be defined as follows.
As shown in Fig.~\ref{fig:stringsAndFlux}, an empty edge carries one unit of magnetic field, oriented from the even to the odd sublattice.
And each edge occupied by a dimer carries two field units going from the odd to the even sublattice. We shall count the magnetic {\em flux}
going through oriented closed paths, the smallest one being a small circle surrounding a single site of the hexagonal lattice. Due to the constraint that every site  is reached by exactly one dimer,
the flux going through such a small circle vanishes -- the lattice divergence of the magnetic field is zero at each site (${\rm div} \vec B=0$). 
Correspondingly, the flux through any contractible loop, such as $\mathcal{C}$ in Fig.~\ref{fig:stringsAndFlux}, also vanishes. Indeed, since ${\rm div} \vec B=0$, one can deform the loop until it reduces to the smallest one. On the other hand, the flux through non-contractible loops can be non-zero. Again, any local dimer rearrangement, including the flip term of the Hamiltonian, conserves the flux through such loops. Since, for the torus, there exist two independent non-contractible loops, this leads to a pair of fluxes $F_x$ and $F_y$ characterizing flip-disconnected topological sectors. The star and plaquette crystals, discussed in the main text, are found in the zero-flux sector, while the staggered dimer configurations are found in the maximal flux sectors, each consisting of a single  (isolated) configuration.

\Section{String representation and fluxes}

Dimer coverings can equivalently be represented as configurations of nonintersecting strings that form closed loops on the lattice
\cite{orland_exact_1993,schlittler_phase_2015}. Specifically, one can start from the staggered dimer configuration depicted on the left of Fig.~\ref{fig:stringsAndFlux} (horizontal dimers only) and choose a certain configuration of closed nonintersecting strings such that, on the path of every string, empty and occupied edges alternate. Now, the corresponding dimer configuration is obtained by doing so-called loop updates along the string paths. These consist in exchanging, on the chosen paths, empty and occupied edges. The winding numbers of these strings are in direct correspondence with the flux quantum numbers $(F_x,F_y)$ introduced above. If we consider, for example, the case with $F_x=0$ that is the relevant sector for the main part of the paper, every dimer covering with flux $F_y$ corresponds to a configuration of $N_s$ strings that encircle the torus in $x$ direction such that
\begin{equation}
	F_y = 2L_y-3N_s\quad\Leftrightarrow\quad
	f = F_y/L_y=2-3N_s/L_y.
\end{equation}

\Section{Height representation}
\begin{figure}[h]
	(a)\hspace{0.28\columnwidth}(b)\hspace{0.28\columnwidth}(c)\\[0.3em]
	{ \includegraphics[width=0.95\columnwidth]{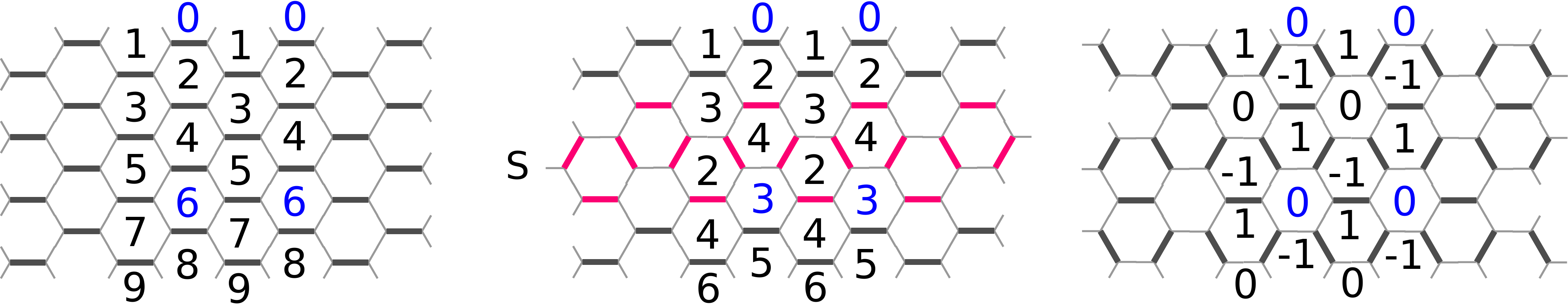} }\\[0.7em]
 \caption{ Height mapping for the hexagonal dimer problem. An integer height is associated with each plaquette according to the rule given in the text. (a) A staggered dimer configuration corresponds to a maximal slope, and therefore the maximal flux sector. (b) Strings (in this case an $S$-string) reduce the slope. (c) In the height representation, the star crystal has (on average) zero slope.}
 \label{fig:height}
\end{figure}
Dimer coverings of a bipartite lattice
can equivalently be 
represented using an integer height associated with each plaquette (dual lattice sites) \cite{blote_roughening_1982,moessner_quantum_2008}. On the hexagonal lattice, turning clockwise around a site of the  even sublattice, the height $h(\vec{x})$ changes by +1 when crossing an empty edge, and  by -2 when crossing a dimer (respectively -1 and +2 around a site of the odd sublattice), as depicted in Fig.~\ref{fig:height}. The magnetic field $\vec B$, defined above, is perpendicular to the slope in this height representation. So, configurations
configurations with zero flux have a vanishing average slope in the height
language (``flat'' configurations). Similarly, configurations with a large flux correspond to a large slope. This maps the dimer covering problems onto faceting problems for surfaces in three dimensions \cite{levitov_equivalence_1990}.
It is indeed conventional to represent dimer coverings on the hexagonal lattice as ``stacks of cubes'' (see for instance Fig.~1b in \cite{levitov_equivalence_1990}).
The height $h$ introduced above then corresponds to the position of the surface of the cubes, after projection onto the (1,1,1) axis of the underlying cubic lattice.
In this language, staggered dimer regions appear as smooth (but tilted) surfaces, normal to the (1,0,0), (0,1,0) or (0,0,1) direction.
On the other hand, (static) $S$-strings at distance $d$
from each other correspond to  steps separating (tilted) terraces of width $d$.
The star crystal corresponds to a microscopically corrugated surface with
a vanishing average slope.

\Section{Dimer sum rules}

Dimer coverings on a tiling by definition satisfy a simple constraint -- each vertex is reached by exactly one dimer. Hence, dimer coverings are constrained by simple sum rules, associated with Euler-Poincar\'e and Gauss-Bonnet relations for tilings on compact surfaces \cite{sadoc_frustration_1999}. 
For a given covering, let $N_{d}$ denote the total number of dimers, $n_j$ the number of plaquettes covered with $j$ dimers, and $N$ and $V$ the total number of plaquettes and vertices, respectively. Calling $j_{\rm max}$ the maximum number of dimers that can sit on a plaquette, this gives the two sum rules
\begin{equation}
	\sum_{j=0}^{j_{\rm max}} j \cdot n_j =2N_d=V\quad\text{and}\quad
	\sum_{j=0}^{j_{\rm max}}n_j =N.
\end{equation}
The first follows from the local dimer constraint which implies that $V=2N_{d}$. The second rule simply expresses that the total number of plaquettes is $N$.
For a regular tiling, i.e., a tiling with constant coordination number $c$,
the numbers of plaquettes and vertices obey $N=V(c/2-1)$ on a torus.

For the hexagonal lattice, we have $c=3$, $j_{\rm max}=3$, and $V=2N$, leading to
\begin{equation}
2n_0+n_1-n_3=0 \quad\text{and}\quad \sum_{j=0}^{3}n_j = N. \label{eq:sumrule}
\end{equation}
Note that, on average, plaquettes carry two dimers. 
For the square lattice, $c=4$, $j_{\rm max}=2$, and $V=N$, leading to the $n_2=n_0$ result noted in the conclusion. Note finally that tilings with fixed boundaries can also be analyzed along the same line, by properly entering additional boundary terms.

\Section{Classical phase diagram}
Let us compute the classical ground states ($t=0$) for arbitrary values of $v_0$ and $v_3$. To determine the  $j$-plaquette densities $\vec{\rho}=(\rho_0,\rho_1,\rho_2,\rho_3)$ that minimize the energy, we first introduce a parametrization of the accessible phase space. According to the sum rules
\begin{equation}
	2\rho_0+\rho_1-\rho_3=0 \quad\text{and}\quad \sum_{j=0}^{3}\rho_j =1,\label{eq:classical}
\end{equation}
the physical states are restricted to a triangular region in the $(\rho_0,\rho_1,\rho_3)$-space, formed by the origin $O =(0,0,0)$ and the points $A=(1/3,0,2/3)$ and $B=(0,1/2,1/2$). We parametrize a generic point $P$ inside that triangle by
\begin{equation}
	P =\left(\frac s3,\frac {1-s}{2},\frac 12+\frac s6 \right)r \quad {\rm with} \quad (r,s) \in [0,1]. \label{eq:classical2} 
\end{equation}

Now, with $v_0=\sin \alpha$ and $v_3=\cos\alpha$, the energy per plaquette $E(\alpha)$ reads \begin{eqnarray}
E(\alpha) &=&\rho_0 v_0+\rho_3 v_3 ,\label{eq:energyclassical} \\
 &=&\left[ \frac s3 \left(\sin\alpha +\frac {\cos \alpha}{2} \right) +\frac 12 \cos \alpha\right]r.
\end{eqnarray}
We now seek for the minimum of $E(\alpha)$ in terms of $r$ and $s$. Clearly, the sign of $\left(\sin\alpha +\frac {\cos \alpha}{2} \right) $ decides whether $s=0$ or $1$ for the ground state configuration. With $\alpha_1=\arctan(-2)\simeq -63.4^\circ$, $\alpha_2=\pi/2-\alpha_1$ and clockwise rotation, one obtains the three regions given in the main text: 

(i) $\alpha\in\left[\pi/2,\alpha_1\right]$: $E(\alpha)\geq 0$ and the ground state energy is minimized (and vanishes) throughout this interval when $r=0$, leading to \mbox{$\vec{\rho}=(0,0,1,0)$}. This corresponds to the staggered states (nonflippable configurations) in the maximal $f=2$ flux sector. When $\alpha=\pi/2$, any configuration satisfying $s=0$, hence the full segment $OB$, also defines a ground state. Such configurations can be found in every flux sector. At the opposite end of this angular sector, when $\alpha=\alpha_1$, all configurations falling in the segment $OA$, therefore $s=1$ and $r\in [0,1]$, also have a vanishing energy. Again, such configurations exist in every flux sector.

(ii)  $\alpha\in\left[\alpha_1,\alpha_2\right]$: The threefold degenerate ground state is the star crystal, corresponding to the point $A$ with \mbox{$\vec{\rho}=(1/3,0,0,2/3)$}. It belongs to the $f=0$ sector (Fig.~1, main text). When $\alpha=\alpha_2$, all configurations falling in the segment $AB$, with $r=1$ and  $s\in [0,1]$, have minimal energy. Such configurations can be found at least in sectors $f\in[0,1/2]$.

(iii) $\alpha\in\left[\alpha_2,\pi/2\right]$: The ground state is a 12-fold degenerate crystalline state, denoted by  $S_2$, with \mbox{$\vec{\rho}=(0,1/2,0,1/2)$}. It belongs to the $f=1/2$ sector. 

For this classical limit, the locations of ground states in the $(\rho_0,\rho_1,\rho_3)$-space show a nice ``dual" relation with respect to the circle (angle $\alpha$) that parametrizes the Hamiltonian. In the $(v_0,v_3)$-plane, the classical phase diagram has three angular sectors separated by the values $(\pi/2,\alpha_1, \alpha_2)$. The set of classical configurations defines a convex region bounded by the triangle $(O,A,B)$ in the $(\rho_0,\rho_1,\rho_3)$-space. As described above, the ground states lie on the boundary of that triangle. Therefore, the continuous $\alpha$ intervals map onto the triangle's vertices, while the three singular values of $\alpha$ are mapped onto the triangle's edges.

\Section{Details about Monte Carlo simulations}

The numerical method used in this work has been detailed in Ref.~\cite{schlittler_phase_2015}, and we therefore only briefly summarize it here. As done by Moessner, Sondhi, and Chandra \cite{moessner_phase_2001}, the 2D quantum dimer model on a hexagonal lattice can be studied by first mapping it to an antiferromagnetic 2D quantum Ising model on the (dual) triangular lattice, comprising  diagonal six-spin interactions.
The resulting  model can be studied efficiently using world-line quantum Monte Carlo \cite{suzuki_monte_1977} by approximating its partition function and observables by those of a classical 3D Ising-type model on a stack of triangular 2D lattices
(quantum-classical mapping). We speed up the Monte Carlo simulation of the classical 3D model through suitable cluster updates.

The equivalence between dimer and spin models is a delicate issue for two 
reasons.
First, as we are free to choose the orientation of some reference spin, a given dimer configuration corresponds to two spin configurations that differ by a global spin flip.
Dimer configurations therefore correspond to the
spin-flip symmetric sector of the spin model. One can nevertheless simulate 
the full spin model in the Monte Carlo as, based on the Perron-Frobenius 
theorem, it can be shown that the global ground state is always in the spin-flip symmetric sector.
Second, half of the topological sectors of the dimer configurations correspond to periodic boundary conditions for the spins,
and half of the sectors correspond to anti-periodic boundary conditions in the Ising model.

\begin{figure}[t]
 \includegraphics[width=8cm]{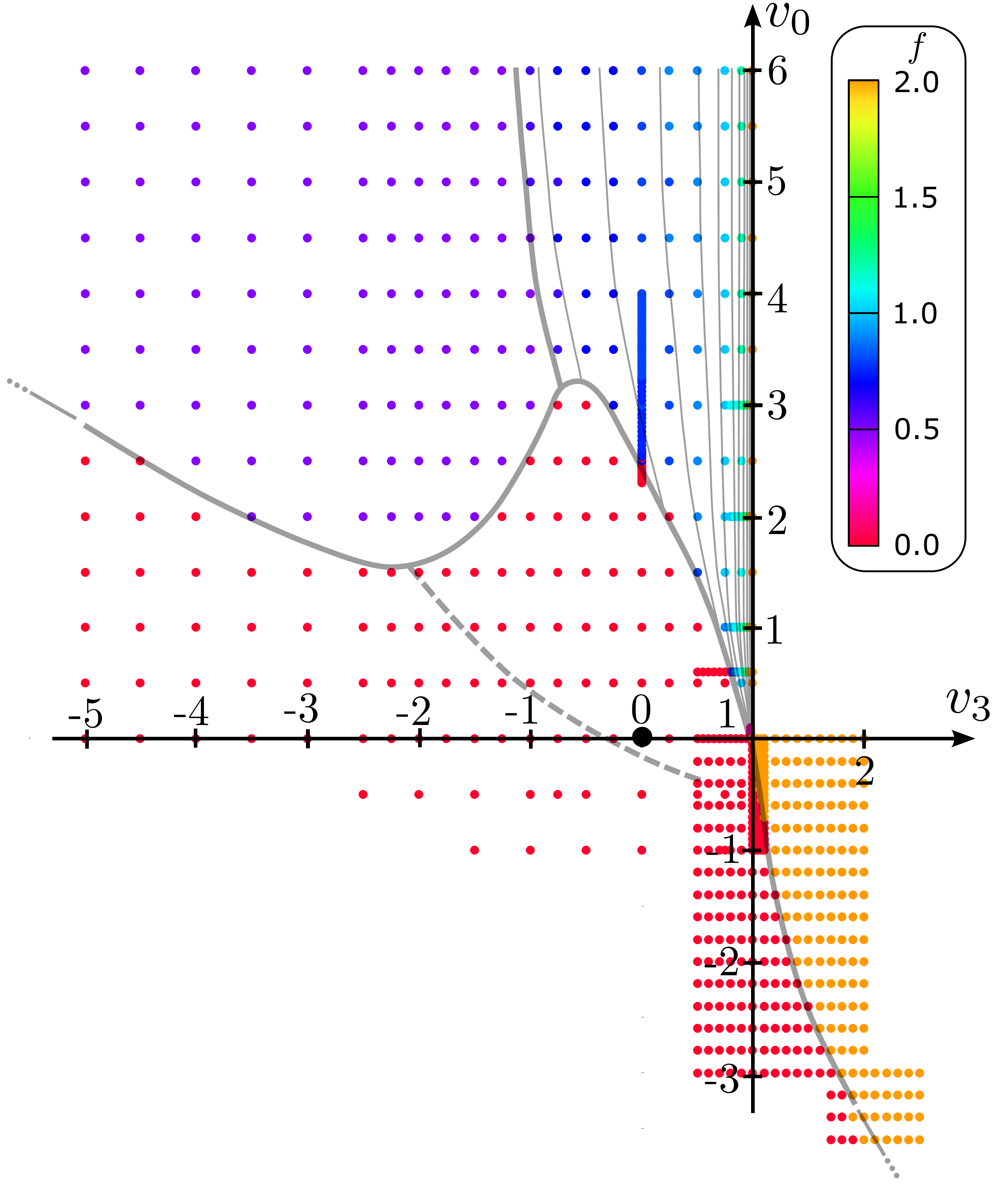}
 \caption{
 Points in the $v_0-v_3$ plane where the simulations were carried out.
 The corresponding ground state fluxes (see color scale) were used to determine the flux transition lines shown here (grey lines) and in Fig.~1 (main text).}
\label{fig:flux_diagram_grid}
\end{figure}

In the present work, we have only used periodic boundary conditions for the Ising spins, and systematically compared the available flux sectors in one direction (say $F_y$), while the other ($F_x$) is kept zero,  by explicitly constructing an appropriate initial spin configuration.
In doing so, we have assumed that states in each flux sector are flip 
connected. It is indeed generally believed, that the local dynamics are 
ergodic in each topological sector, besides those of maximum flux (see 
Prop.~2.3 in Ref.~\cite{propp_lattice_2002}).
Most of the simulations were carried out for a rectangular cluster with
$60\times60$ plaquettes, as noted in the main text, allowing for 20 different
flux densities $f_y$, equally spaced by steps of $\delta f = 0.1$.
The points in the $v_0-v_3$ plane that were investigated to determine the flux transition lines in Fig.~1 (main text) are displayed in Fig.~\ref{fig:flux_diagram_grid}.
Other system sizes, ranging from $56 \times 55$ to $120 \times 120$ plaquettes, were also used,
mainly to study in more detail the transitions along the  $v_3 = 0$ line
(corresponding to the cluster of points for $2.5 \leq v_0 \leq 4$ in Fig.~\ref{fig:flux_diagram_grid}). We verified
in particular that the energy density is the same for two systems with different sizes
but same flux density.

For the QDM at $v_0=0$, the transition between the star and plaquette 
phase occurs in the zero-flux sector and can be detected using magnetization 
variances of the corresponding spin model 
\cite{moessner_phase_2001,schlittler_phase_2015}. In the more generic case 
with nonzero $v_0$, studied in this work, many transitions between {\em different} flux sectors occur.
These must therefore be detected by comparing the energies of different sectors. As detailed in Ref.~\cite{schlittler_phase_2015}, the energy can be evaluated using imaginary-time spin-spin correlators $\langle\sigma_i^n\sigma_i^{n+1}\rangle$. We scanned the $(v_0,v_3)$ plane in order to determine the ground state flux.
In addition, the expectation values of several spin and dimer observables (dimer densities and correlations)
were measured to characterize each phase. In particular, plots of the average dimer  occupancies (for each plaquette) are used to visualize the spatial/crystalline structures and compare stripe-like organizations in the ``fan'' region.

\Section{Perturbation around the RK point}

\subsection{Classical transfer matrix and free fermions}

In this section we compute the dimer density expectation values
$\rho_0$ and $\rho_3$ in the 
RK ground state, as a function of the flux density $f$.
RK states being equal amplitude superpositions of all covering in a given flux sector,
$\rho_0$ and $\rho_3$ are also the densities of a classical statistical 
dimer problem at infinite temperature. We solve the later using a transfer matrix method.

\begin{figure}[h]
 \includegraphics[width=5cm]{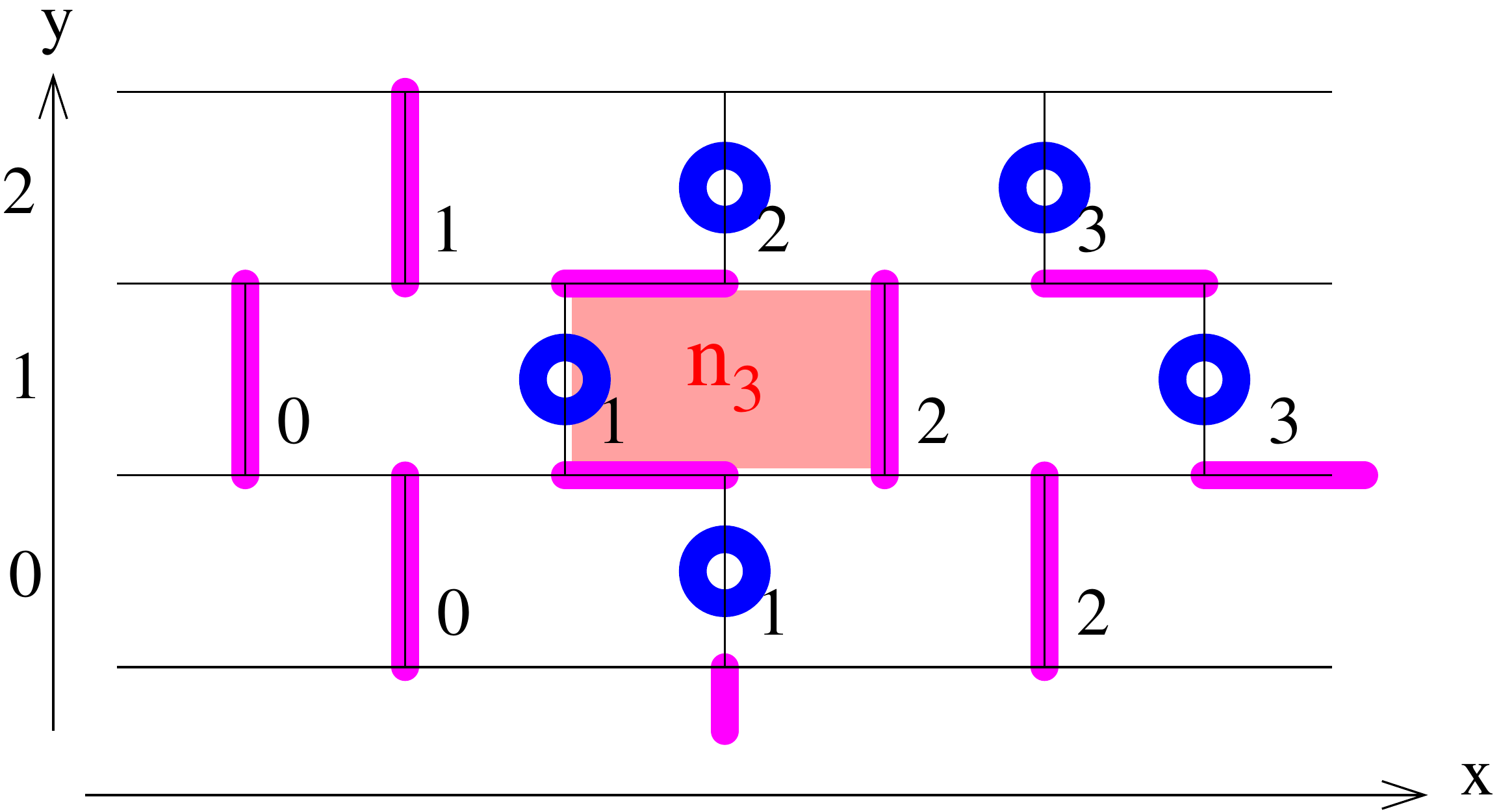}
 \caption{Brickwall representation of the hexagonal lattice. The fermions (blue circles) live on the vertical edges
 which are {\em not} occupied by a dimer (magenta). The transfer matrix propagates the configurations
 in the positive $y$ direction from row to row.
  Note the numbering of the ``sites'' (vertical edges):
 a fermion on site $x$ may go to $x$ or $x+1$ in the subsequent row. A \mbox{$3$-plaquette} is shaded.  To enforce the presence of three dimers around this plaquette we need:
 (i) row $y=0$: one fermion on edge 1;
 (ii) row $y=1$: one fermion on edge 1 and one hole on edge 2 (or vice versa);
 (iii) row $y=2$:  one fermion on edge 2. The Eq.~\eqref{eq:n3} is the associated expectation value.
 }
 \label{fig:TM}
\end{figure}
\begin{figure}[h]
 \includegraphics[width=5cm]{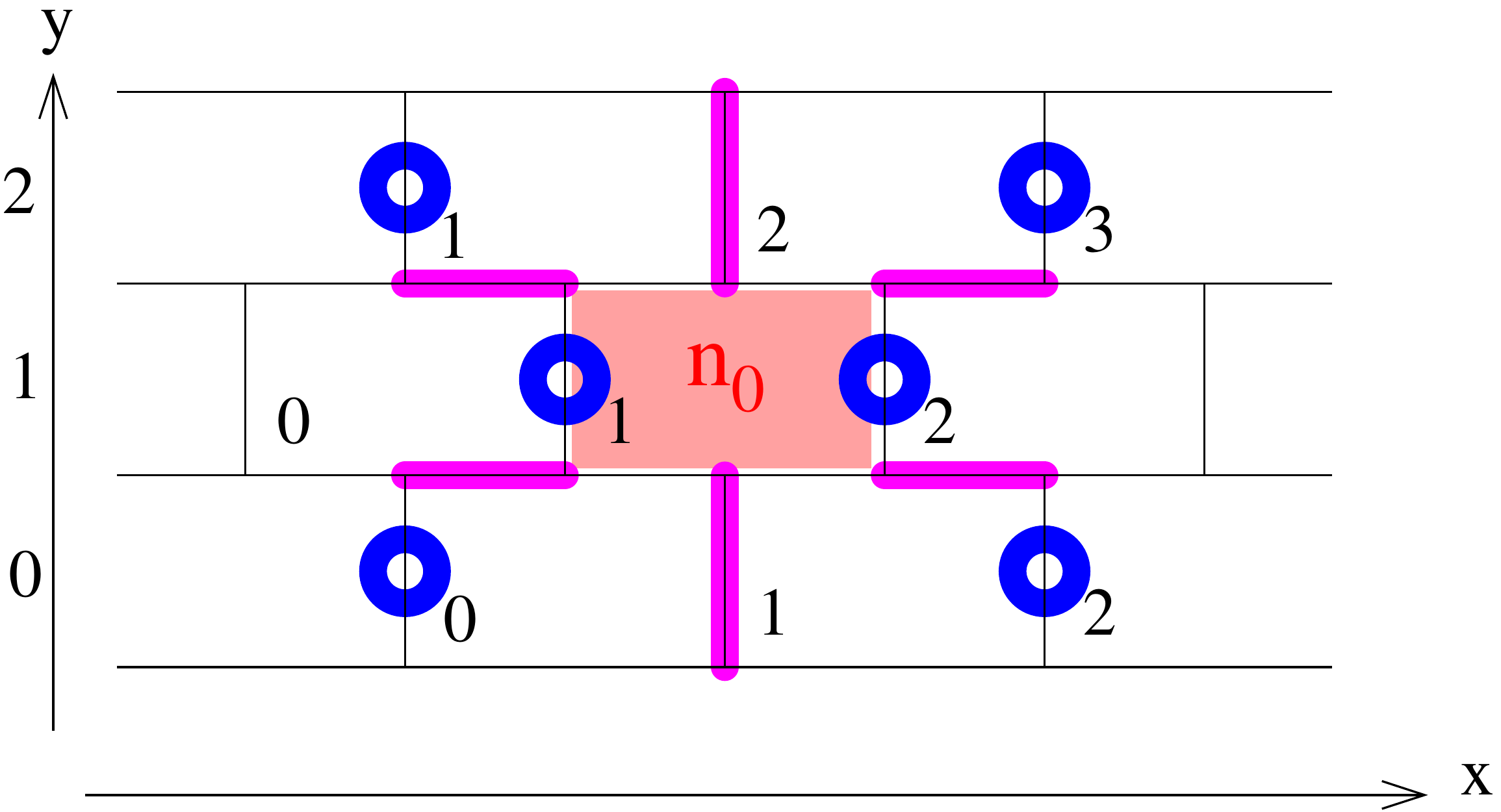}
 \caption{Same as in Fig.~\ref{fig:TM}, but now for a configuration with a \mbox{$0$-plaquette} (shaded). To have this 0-plaquette, in terms of the 
fermionic representation, we need
 (i) row $y=0$: a hole on edge 1 (ii) row $y=1$: two fermions on the edges 1 and 2, and (iii) row $y=2$: a hole on edge 2.
 See Eq.~\eqref{eq:n0}.
 }
 \label{fig:TM_n0}
\end{figure}

To treat the classical dimer model on the hexagonal lattice using a transfer matrix it is convenient to consider
the ``brick wall'' version of the lattice, as depicted in Fig.~\ref{fig:TM}.
First, note that it is sufficient to consider the dimer occupations of the vertical edges -- the information on all the other edges can be obtained using the hard-core constraints. 
The transfer matrix $\hat T$ then relates a dimer configuration $|\psi\rangle$ on one row $y$ to the
configuration on the row $y+1$ above. More precisely, $\hat{T}|\psi\rangle$ is the linear superposition of
all the configurations of the row $(y+1)$ which are compatible with $|\psi\rangle$ at level $y$.
The next step is to consider a single row of vertical edges, and to associate a (spinless) fermion Fock space to it:
 edges not occupied by a dimer carry the fermions, and  edges with a dimer carry (fermionic) holes.
In particular, the Pauli exclusion principle enforces the dimer hard-core constraint.
We also note that the fermions (and their world lines) correspond exactly to the strings discussed  in the main text.
The $x$ component of the flux density is simply related to density of vertical dimers, which is, in turn, simply related to the fermion density $n$ (which is the same for each row):
\begin{equation}
 f=2-3n. \label{eq:f_fermions}
\end{equation}
Thus, the particle number conservation of the transfer matrix enforces the 
flux conservation.
If we note $c^\dagger_x$ the fermion creation operator
on site $x$ (see the numbering in Fig.~\ref{fig:TM}), the transfer matrix can be shown to obey:
\begin{eqnarray}
 \hat{T} c^\dagger_x &=& \left(c^\dagger_x + c^\dagger_{x+1}\right) \hat{T} ,\label{eq:Tc} \\
\hat{T} |{\rm vacuum}\rangle &=&|{\rm vacuum}\rangle .
\end{eqnarray}
In other words, a fermion on site $x$ should propagate  to $x$ or $(x+1)$ in the line above. 
Performing a Fourier transform of Eq.~\ref{eq:Tc} gives
\begin{eqnarray}
 \hat{T} c^\dagger_k &=& c^\dagger_k \left(1  + e^{ik}\right) \hat{T} \label{eq:Tc_k},
\end{eqnarray}
where $c^\dagger_k$ is the Fourier transform of $c^\dagger_x$.
This shows that $\hat{T}$ is a product of operators acting independently on each Fourier mode.
The solution is \cite{stephan_shannon_2009}:
\begin{eqnarray}
 \hat{T} =\prod_{k\in [-\pi,\pi[} \left( 1+e^{ik} c^\dagger_k c_k \right).
\end{eqnarray}
From this one can find the commutation relations with annihilation operators:
\begin{eqnarray}
 c_k \hat{T}  &=&  \hat{T} c_k \left( 1 + e^{ik} \right), \label{eq:cT_k} \\
 c_x \hat{T}  &=&  \hat{T} \left( c_x + c_{x-1} \right)\label{eq:cT}.
\end{eqnarray}
In the following it will also be necessary to commute $c_x$ (and $c^\dagger_x$) and $\hat{T}$ in the reversed direction compared to Eqs.~\eqref{eq:Tc}
and \eqref{eq:cT}. The results are now infinite sums:
\begin{eqnarray}
  c^\dagger_x  \hat{T} &=& \hat{T} \left(c^\dagger_x - c^\dagger_{x+1} + c^\dagger_{x+2} - c^\dagger_{x+3}+\cdots\right) , \label{eq:cT2} \\
 \hat{T}c_x  &=&   \left(c_x-c_{x-1}+c_{x-2}-c_{x-3}+\cdots \right)\hat{T}. \label{eq:Tc2}
\end{eqnarray}

When the $y$ dimension of the lattice goes to infinity, only the eigenvector of $\hat{T}$ with the largest eigenvalue in the given flux sector needs to be kept.
The later is nothing but a Fermi sea $|f\rangle$ with Fermi momentum $k_F$ and density $n=k_F/\pi$.
The corresponding eigenvalue, $\Lambda(k_F)=\prod_{-k_F<k\leq k_F} (1+e^{ik})$, allows to compute the entropy per site, but its explicit expression is not needed here.

\subsection{Density of \texorpdfstring{$3$}{3}-plaquettes}
We start by the computation of $\rho_3(f)$, the density of $3$-plaquettes. A corresponding hexagon is shaded in Fig.~\ref{fig:TM},
and it is characterized by one fermion in $x=1$ on the lowest row (thus associated with the projector $c_1^\dagger c_1$), one fermion in $x=1$ and one hole in $x=2$ 
in the second row ($\to c_1^\dagger c_1 c_2 c_2^\dagger$), and, finally, a fermion at $x=2$ in the third row ($\to c_2^\dagger c_2$).
The density $\rho_3$ is thus
\begin{equation}
 \rho_3=2 \frac{ \langle f |  c_2^\dagger c_2 \hat{T} c_1^\dagger c_1 c_2 c_2^\dagger \hat{T} c_1^\dagger c_1  | f \rangle}
 {\langle f | \hat{T}^2 | f \rangle},
  \label{eq:n3}
\end{equation}
(the factor 2 is due to the fact that there are two ways to put three dimers around an hexagon).
The next step amounts to eliminate $\hat{T}$ by using the relations Eqs.~\eqref{eq:Tc},\eqref{eq:cT},\eqref{eq:cT2} and \eqref{eq:Tc2}.
The result is
\begin{equation}
 \rho_3= 2 \langle f| \hat{D}^\dag_2(c_2+c_1)c_1c_1^\dag c_2^\dag c_2(c^\dag_1+c^\dag_2)\hat{S}_1| f \rangle,
 \label{eq:r3A}
\end{equation}
where we have defined:
\begin{eqnarray}
 \hat{D}^\dag_r&=&\sum_{x=0}^\infty (-1)^x c^\dag_{x+r},\\
 \hat{S}_r&=&\sum_{x=-\infty}^0 (-1)^x c_{x+r}.
\end{eqnarray}
The correlator of Eq.~\eqref{eq:r3A} can be obtained, using Wick's theorem, as the determinant of a $4\times4$ matrix $M$:
\begin{widetext}
\begin{eqnarray}
 M_3&=&\left(
 \begin{array}{cccc}
 \langle \hat{D}^\dag_2 (c_2+c_1) \rangle  &  \langle \hat{D}^\dag_2 c_1 \rangle  & \langle \hat{D}^\dag_2 c_2 \rangle & \langle \hat{D}^\dag_2 \hat{S}_1 \rangle \\
-\langle (c_2 +c_1) c^\dag_1 \rangle & -\langle c_1 c^\dag_1 \rangle  & \langle c^\dag_1 c_2 \rangle & \langle  \hat{S}_1  c^\dag_1 \rangle \\
-\langle (c_2 +c_1) c^\dag_2 \rangle & -\langle c_1 c^\dag_2 \rangle  & \langle c^\dag_2 c_2 \rangle & \langle c^\dag_2 \hat{S}_1 \rangle \\
-\langle (c_2 +c_1) (c^\dag_1+c^\dag_2) \rangle & -\langle c_1  (c^\dag_1+c^\dag_2) \rangle  & -\langle  c_2  (c^\dag_1+c^\dag_2)\rangle & \langle   (c^\dag_1+c^\dag_2) \hat{S}_1 \rangle
 \end{array}
  \right).
  \end{eqnarray}
\end{widetext}
The two-point functions appearing above can be expressed using the 
  correlator of the Fermi sea:
  \mbox{$G_{x-y}=\langle c^\dag_x c_y \rangle = \frac{\sin[n\pi(x-y)]}{\pi(x-y)}$} for $x\ne y$, and $\langle c^\dag_x c_x \rangle=n$.
The correlations  $\langle \hat{D}^\dag_x c_y\rangle$ or $\langle c_x \hat{S}^\dag_y \rangle$ contain some infinite sums which can be evaluated
using the sum rule: \mbox{$\sum_{r=0}^\infty  (-1)^r G_r = n/2$}.
The one appearing in $M_3$ are $\langle \hat{D}^\dag_2 c_1\rangle =\langle \hat{D}^\dag_2 c_2\rangle=\langle c^\dag_2 \hat{S}_1 \rangle=\langle c^\dag_1 \hat{S}_1 \rangle=n/2$.
The last one, $\langle \hat{D}^\dag_2 \hat{S}_1 \rangle$, contains two sums which can also be performed exactly, leading to 
$\langle \hat{D}^\dag_2 \hat{S}_1 \rangle=\frac {\sin ( n\pi) }{2\pi \left[ 1+\cos ( n\pi) \right] }$.
The matrix $M_3$ therefore takes the explicit form:
\begin{eqnarray}
  M_3&=&\left(\begin {array}{cccc}
  n&n/2 &n/2&\frac {\sin \left( n\pi  \right) }{2\pi \left[ 1+\cos \left( n\pi  \right)  \right] }\\
  A&n-1& G_1 &n/2\\
  A&G_1&n&n/2\\
  2A&A&A&n
  \end {array} \right),
\end{eqnarray}
where $G_1=\frac{\sin(\pi n)}{\pi}$ and  we have set $A=G_1-1+n$.
The quantity $\rho_3$ is finally obtained from the determinant of $M_3$:
\begin{eqnarray}
 \rho_3 &=& 2 \det(M_3) ,\nonumber\\
 &=& \frac { \left(  \left[ 2+\cos \left( n\pi  \right)  \right] {n}^{2}-2n+1 \right) \sin \left( n\pi  \right) }{\pi \, \left[ \cos \left( n\pi  \right) +1 \right] }\nonumber\\
 &&-{n}^{2} \left( n-1 \right) +{\frac {\sin \left( n\pi  \right)  \left[ \cos \left( n\pi  \right) -1 \right] }{{\pi }^{3}}}.
\end{eqnarray} 

\subsection{Density of \texorpdfstring{$0$}{0}-plaquettes}
The density of $0$-plaquette can be obtained in a similar way. The starting point is the following correlator (see Fig.~\ref{fig:TM_n0}):
\begin{equation}
 \rho_0= \frac{ \langle f |  c_2 c_2^\dagger \hat{T} c_1^\dagger c_1 c_2^\dagger c_2 \hat{T} c_1 c_1^\dagger  | f \rangle}
 {\langle f | \hat{T}^2 | f \rangle}.
 \label{eq:n0}
\end{equation}
After commutating one $\hat{T}$ to the right and the other to the left we get:
\begin{equation}
 \rho_0= 2 \langle f| (c_2+c_1) \hat{D}^\dag_2 c_1^\dag c_1 c_2^\dag c_2\hat{S}_1 (c^\dag_1+c^\dag_2)| f \rangle.
\end{equation}
As for $\rho_3$, we construct a matrix from the two-point contractions and the result is:
\begin{eqnarray}
M_0=\left( \begin {array}{cccc}
n-1 & n/2 & n/2 &{\frac {\sin \left( n\pi  \right) }{2\pi \left( \cos \left( n\pi  \right) +1 \right) }}\\
A & n &  G_1 & n/2\\
A &G_1 	&n&n/2\\
2A & A & A &n-1
\end {array} \right).
\end{eqnarray} 
Finally $\rho_0$ is obtained by taking the determinant:
\begin{eqnarray}
 \rho_0&=& \det(M_0),\nonumber \\
 &=&\frac {\cos \left( n\pi  \right)  \left( \cos \left( n\pi  \right) +1 \right) +{n}^{2}{\pi }^{2} \left( n-1 \right) -2}{{\pi }^{2}}\nonumber\\
 &&-n \sin \left( n\pi  \right)
    \frac {   \cos \left( n\pi  \right)  \left( n-2 \right) +2n-3  }
    {{\pi } \left( \cos \left( n\pi  \right) +1 \right) }\nonumber \\
 &&-\frac{1}{\pi^3}\sin \left( n\pi  \right) \left(\cos \left( n\pi  \right)-1\right).
\end{eqnarray}

Note that we also computed $\rho_1$ using the same method and we checked that the  sum rule $\rho_3=\rho_1+2\rho_0$ is satisfied.

\Section{Some variational states based on simple string arrangements}

\begin{figure*}[t]
 \includegraphics [width=\linewidth]{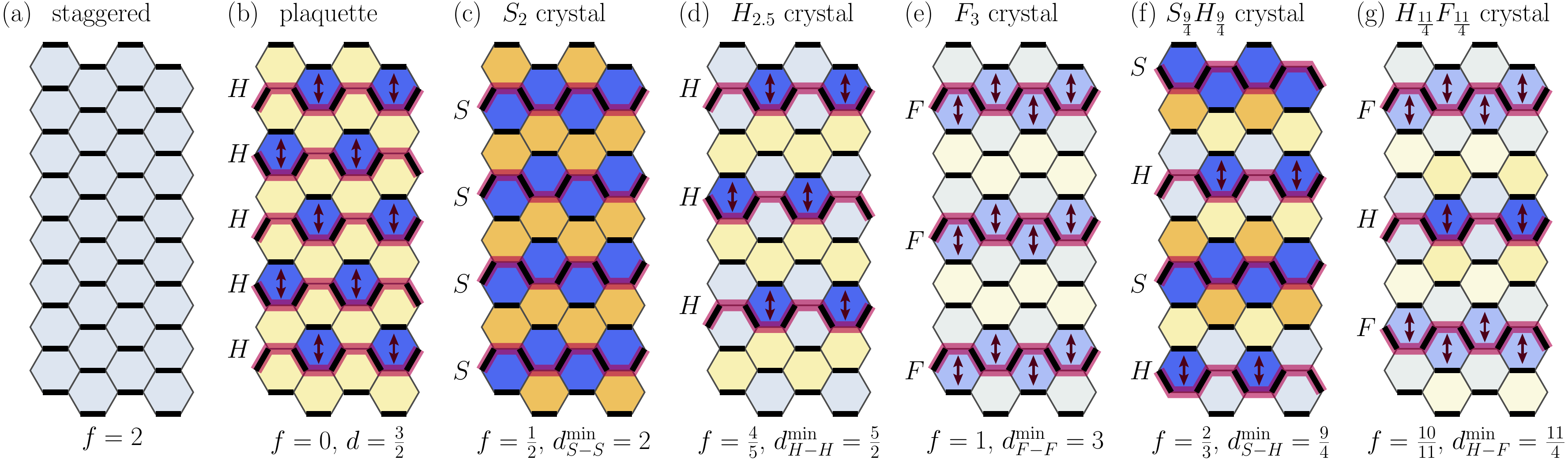}
 \caption{
 Some variational states constructed by stacking 
$S$-, $H$-, and $F$-strings. (a) The staggered state which has flux density 
$f=2$ and corresponds to zero strings. (b) The ideal plaquette state with 
$f=0$, where all plaquettes of one of the three triangular sublattices are in 
the resonating state $\left|\protect\ha\right>+\left|\protect\hb \right>$. This corresponds to $H$-strings at distance 
$3/2$. (c)-(g) Different variational states built as tensor products of $S$-, 
$H$-, and $F$-string states in order to approximate certain crystals in the 
fan region for large $v_0$. In these examples, the strings are at their 
minimal relative distances, such that the states are free of 0-plaquettes. The
plaquette shading corresponds  to the average number of dimers per hexagon, 
using the same color scale as in Fig.~3 of the main text (and in Fig.~\ref{fig:stringsAndFlux}). In
particular, 3-plaquettes are colored in dark blue, 2-plaquettes in light gray-blue, and 1-plaquettes in yellow.
}
\label{fig:var_states}
\end{figure*}

In the main text we mentioned that some density patterns
observed in the QMC simulations can be described by combining some particular  ``building blocks'', called $S$-, $H$- and $F$-strings.
Below we elaborate on this idea.

An $S$-string is a static zigzag configuration, as depicted in Fig.~2 (main text). It corresponds 
to a row of 3-plaquettes.
As an example, the classical star  crystal can be viewed as a periodic arrangement of such \mbox{$S$-strings} at an average distance of $1.5$ (measured in units of the distance between two nearest hexagon centers), noted therefore as a ``classical $S_{1.5}$ crystal".
$H$-strings are set of configurations where, starting from a zigzag configuration, the string can move up by one on every second column. As for $F$-strings, they can move up by one on 
every column.

Let us first suppose that we start with an isolated $S$-string, surrounded only by plaquettes carrying less that $3$ dimers. Switching on the kinetic term of the Hamiltonian, a $3$-plaquette located along the $S$-string is allowed to flip, with the constraint that none of its neighboring plaquettes has already flipped (the condition for still being a $3$-plaquette). We dub this  constrained quantum system  ``$F$-string''. Note that, upon iterated flips, new $3$-plaquettes can be generated which were not bounded by the initial $S$-string, increasing the lateral extension of the set of $3$-plaquettes. Such string configurations will play a  role in high flux sectors, 
which have low string densities,  but we do not consider further these extended chains in this present description. 

These constrained $F$-strings are interesting objects for themselves (see below). But their main interest here comes while considering their regular arrangements a distance $d$: indeed, for $d<3$, correlated flips on two neighboring such strings may create $0$-plaquettes in between, leading to an energy cost in the $v_0>0$ part of the phase diagram.  The QMC simulations show regions where the ground state
dimer densities display such linear zigzag arrays of $3$-plaquettes, regularly spaced at a distance $d$, which we therefore call $S_d$ and $F_d$ crystals.
Now, other nearby regions in the phase diagram (again in the ``fan" region) show different patterns, such that only one over two of the $3$-plaquettes is found to flip significantly. We call these configurations $H$-strings ($H$ for ``half"), and their associated regular arrangements the $H_d$  crystals. Note that, being second neighbours in the zigzag chain, these plaquettes are free to flip, and we can already consider these \mbox{$H$-strings} as having a simple resonating nature. An interesting feature is that the condition for not generating interstring $0$-plaquettes is now that $d\geq d_{H-H}^{\rm min}=5/2$, instead of the above $d\geq d_{F-F}^{\rm min}=3$ for the $F$-strings. We therefore face an interesting competition between $F_d$ and $H_d$  crystals: the $F$-string has potentially a lower kinetic energy, but their dynamics can generate $0$-plaquettes at a shorter interstring  distance (as compared to the $H_d$ crystals).  
 
When $v_0$ gets large (and positive) we find that the ground state energy remains negative (even for $v_3\geq 0$), meaning that the system can simultaneously gain some kinetic energy through resonances on $3$-plaquettes while keeping a vanishing $\rho_0$ (in the limit $v_0\to\infty$). Fig.~\ref{fig:var_states} shows some of the $0$-plaquette free states which can be obtained by stacking $S$-, $H$- and $F$-strings. For these flux sectors, the density patterns observed in the simulations approximately match those of these simple ans\"{a}tze.

We now discuss the energies of
the $F_d$ and $H_d$ quantum crystals. 

According to whether $d$ is integer or half integer, neighboring strings have parallel or anti-parallel zigzag configurations. It is also easy to relate this distance $d$ to the flux sector: in unit of the first distance between plaquette centers, one finds $d=3/(2-f)$
(see Eq.~\ref{eq:f_fermions} and note that $n=1/d$).

Let us, again, first consider an isolated \mbox{$H$-string} with independent resonant plaquettes.
One then forms the tensor product of the individual chain ground states, and check whether this state leads to a useful variational approximation. A necessary condition is  that this state  should not contain any $0$-plaquette.
Such plaquettes would appear in between two such chains if $d$ is too short. As said above, this will not occur whenever $d\geq 5/2$ (equivalently $f\geq 0.8$), giving an upper bound for the ground state energy. In the $v_3=0$ case, this leads to  $E(f)\leq -(2-f)/6$, for $f \in [0.8,2]$. The QMC simulations in the $f=0.8$ sector gives an energy $\sim-0.22$, in rough agreement with the approximate $-0.2$ value found for $H_{5/2}$.

Let us now analyze an isolated \mbox{$F$-string}. Start with a configuration where each plaquette carries 3 dimers, and flip one of these. The flipped plaquette is still a $3$-plaquette, and can therefore be later flipped back. But its two neighboring plaquettes now only carry 2 dimers, and are ``frozen". Repeating the flips on another $3$-plaquettes allows one to span the full Hilbert space for such an isolated chain, containing constrained configurations of flippable or frozen plaquettes. For an open chain of length $L$
the associated Hilbert space dimension is the Fibonacci number $F_L$.
For a closed periodic chain the dimension is $F_L+F_{L-1}$.
It is possible to show an exact correspondence with the Hilbert space of
so-called Fibonacci anyonic chain~\cite{feiguin_interacting_2007}.
We numerically studied the quantum dimer Hamiltonian on a single chain and  
obtained for $v_3=0$ a ground state energy per plaquette $\simeq E_{F}=-0.6035605(9)$. The next step consists in building
a tensor product of these states, which avoid $0$-plaquettes. This is possible if $d\geq d^{\rm min}_{F-F}= 3$, which means $f\geq 1$, giving  upper bound $E(f)\leq (2-f)E_{F}/3$. For $f=1$ ($F_3$ crystal), this gives an upper bound $E_{F}/3$ which is slightly lower than that obtained above for the $H_{5/2}$ state.

Back to the numerical results, we found that, for $v_3=0$ and $2.4 \lesssim v_0\lesssim 12$, the ground state belongs to the $f=4/5$ sector,  with a symmetry well described by that of the $H_{2.5}$ crystal.
Note also that the above variational argument suggests that, at large $v_0$, the sector $f=10/11$  would be be close in energy to that of $f=4/5$. The corresponding state is an alternation of $F$- and $H$-strings at average distance $d=2.75$ (see Fig.~\ref{fig:var_states}). We indeed find numerically that the energy in the sector $f=10/11$ falls below that of $f=4/5$ when $v_0\gtrsim 200$, and displays the expected  $H-F$ density pattern.

More complex states appear to be selected for $v_0\gtrsim 12$. Some of these states can be qualitatively understood by introducing correlations among the strings. A simple family of building blocks consists in forming $n$ coupled $F$-strings.
There, plaquette flips are not only constrained by the state of the two neighbouring plaquettes along the string, but also along the perpendicular direction. One then form a tensor product of these blocks of correlated strings to construct a state of the whole system. This leads to the infinite series of fluxes  $f(n)=(4n+2)/(5n+1)$, running from $f=1$ (single strings, $F_3$ crystal) to $f=4/5$ (all strings coupled, the $H_{2.5}$ case).
We numerically found that the flux $f=6/7$ compatible with 4 coupled strings indeed replaces the $f=4/5$ sector as the ground state for $v_0\gtrsim 12$.

\Section{Transition from the $f=1/2$ sector to the fan region}
In the main text, a variational result for the transition line between the $S_2$-crystal ($f=1/2$) and the fan region ($1/2<f<2$) was stated for large $v_0$. We discuss it here in a bit more detail. As an ansatz for the $S_2$ ground state, we employ $|\dots SSSSSSSSSSSS \dots\rangle$, a regular arrangement of static $S$-strings with distance $d_{S-S}=2$ between neighboring strings. It has an energy of $E/N=v_3/2$ per hexagon. We compare it with a state that has an infinitesimally increased flux density (one string less such that $\Delta F_y=3$). For the corresponding ansatz state, some $S$-strings are  replaced by $H$-strings such that 
we change the configuration above to $|\dots SSHSHSHSHSS\dots\rangle$ with (average) string distances $d_{S-S}=2$ and $d_{S-H}=2.25$. To keep the total system size constant size, note that we need to replace five $S$-strings by four $H$-strings. The resulting energy difference per column of hexagons is
\begin{equation*}
	\Delta E/L_x = -5v_3 +4\left(-\frac{t}{2}+3\frac{v_3}{4}\right),
\end{equation*}
which vanishes at $v_3/t=-1$. This shows that a simple $S_2$ crystal gets destabilized with respect to $H$-string insertion for $v_3/t>-1$.
This simple variational argument thus predicts
that the transition to the fan region occurs (for $v_0\to\infty$) at $v_3/t\approx -1$, indeed not far from the observed value.

\emph{Acknowledgments.---}We wish to thank C.~Boutillier for bringing Ref.~\cite{propp_lattice_2002} to our attention.

\end{document}